\documentclass[aip,reprint]{revtex4-1}
\usepackage{graphicx}

\begin{document}

\title{Photoacoustic Tomography using a Michelson Interferometer with Quadrature Phase Detection}

\author{Rory W. Speirs}
\affiliation{School of Physics, The University of Melbourne, Victoria, 3010, Australia}

\author{Alexis I. Bishop}
\email{Alexis.Bishop@monash.edu}
\affiliation{School of Physics, Monash University, Victoria, 3800, Australia}

\date{\today}

\begin{abstract}
We present a pressure sensor based on a Michelson interferometer, for use in photoacoustic tomography. Quadrature phase detection is employed allowing measurement at any point on the mirror surface without having to retune the interferometer, as is typically required by Fabry-Perot type detectors. This opens the door to rapid full surface detection, which is necessary for clinical applications. Theory relating acoustic pressure to detected acoustic particle displacements is used to calculate the detector sensitivity, which is validated with measurement. Proof-of-concept tomographic images of blood vessel phantoms have been taken with sub-millimeter resolution at depths of several millimeters.
\end{abstract}

\maketitle 

Photoacoustic imaging has the potential to become a routinely used medical imaging modality, combining the superior contrast of optical techniques, with the penetration depth of ultrasound.\cite{Minghua2006, Wang2009, Wang2012} Its inherent ability to distinguish regions of contrasting optical absorption make it an ideal candidate for imaging vascular structure, with possible applications in diagnosis of stroke\cite{Joshi2010} and early stage cancers.\cite{Esenaliev1999, Oraevsky2001, Yang2012}

Photoacoustic tomography (PAT) is potentially capable of producing real time, three-dimensional (3D), high resolution images to depths of several centimeters.\cite{Wang2008, Xiang2013} The requirements of the ultrasonic detection system to achieve this are formidable, and as yet, no system has emerged which satisfy all criteria simultaneously.

Though much emphasis has been placed on detector sensitivity, there are other, equally pressing requirements of a high resolution system. These include the need for a large detection surface, with high spatial resolution. Of particular importance for a system that can be used in clinical context is the ability to rapidly capture data, ideally over the whole detector surface simultaneously. A detector which is transparent to the excitation light is also favourable, as this allows a large amount of optical energy to be dumped uniformly on the region being imaged.

Piezoelectric detectors struggle with many of these requirements,\cite{Hou2007} and so a wide variety of optical detectors have been developed.\cite{Beard1996, Hamilton1998, Paltauf2007, Chow2011} Planar Fabry-Perot based systems show good sensitivity and bandwidth response, but are typically slow to acquire data because of the need to tune the probing laser at each point on the detector surface to achieve peak sensitivity.\cite{Zhang2008} Simultaneous two-dimensional (2D) data collection has been demonstrated with these systems,\cite{Lamont2006} but large detection areas are challenging to produce because of the difficulty in creating polymer coatings of uniform thickness.
 
Microring resonators have been made with impressive sensitivity, element size and can be made transparent. However coupling and addressing a large array of microrings will be difficult, so it is yet to be seen if simultaneous full surface detection is achievable.\cite{Huang2008} 

Pressure dependent optical reflectance detectors have been demonstrated with the ability to capture pressure data over a whole surface simultaneously,\cite{Paltauf1999,Paltauf1997} without the need for complicated nanofabrication techniques of some other methods. Moreover, use of fast-framing, or gated charge-coupled devices (CCDs) simplifies data collection, and allows for high spatial resolution over a large detection surface. However, the detection sensitivity of this type of system has so far been only modest, and may be insufficient for high resolution imaging of biological tissue.

We have developed a detector based on a Michelson interferometer (MI) with quadrature phase detection. This detector has comparable sensitivity to other optical detectors in the literature, but has the potential to perform high resolution measurements over a full 2D surface simultaneously, without the need for any position dependent sensitivity tuning. 

The MI acts as an ultrasound sensor simply by acoustically coupling an ultrasound source to a mirror in one of the arms. The acoustic wave of pressure $p$, has an associated particle displacement $\xi$, which shifts the position of the mirror as the wave passes through it. This change in position adjusts the relative phase of the laser beams in the two arms, resulting in a change in fringe brightness at the output of the interferometer. For a small amplitude wave travelling in the $x$ direction at time $t$, pressure and displacement are related by:
\begin{equation}
	p(x,t)=-E\frac{\partial\xi(x,t)}{\partial x},
	\label{eqn:pressure_displacement}
\end{equation}
where $E$ is the appropriate modulus of elasticity for the medium.\cite{Blitz1963}

The intensity, $I$ of the recombined beam in a standard MI varies sinusoidally with mirror position:
\begin{equation}
I=\frac{I_{0}}{2}(1+\cos(\phi)),
\label{eqn:michelsonintensity}
\end{equation}
where $\phi=4\pi n\xi/\lambda$. Here, $I_0$ is the input intensity, $\phi$ is the phase, $\lambda$ is the wavelength of the probe beam, and $n$ is the refractive index of the arm where the mirror position is changing.

Michelson interferometers previously used in ultrasonic detection therefore suffer from the same problem as Fabry-Perot type detectors, in that the laser wavelength (or mirror position) must be tuned to a sensitive region at each point in order to obtain good optical modulation for a given mirror displacement.\cite{Mezrich1976}

The need for tuning was removed from our system by employing quadrature phase detection. In quadrature phase detection, two orthogonal linear polarizations are used to simultaneously obtain two separate interference patterns at the output of the interferometer, which have a relative phase difference of $\pi/2$. This phase difference ensures that the interference pattern of at least one of the polarization components is always in a sensitive region.

The phase shift between the two polarization components is created by first linearly polarizing the light at 45$^\circ$ from the vertical or horizontal axis of the polarizing beamsplitter. A liquid crystal variable waveplate is placed in one of the arms, which retards the phase of one polarization component relative to the other by nominally $\pi/4$ in both the forward and reverse trips. A variable retarder is used instead of a fixed $\lambda/8$ wave plate to compensate for small amounts of birefringence present in other optical components. 

The \emph{phase sensitivity}, $\frac{dI}{d\phi}$ (which is the optical intensity modulation per radian of phase) of  a normal MI varies between $0~\rm rad^{-1}$ and $I_0/2~\rm rad^{-1}$. For an MI with quadrature detection, the phase sensitivity for each polarization is simply added together, so is always between $I_0/2~\rm rad^{-1}$ and $I_0/\sqrt{2}~\rm rad^{-1}$. This ensures that the total sensitivity of the system is always at least as high as the maximum of a standard MI, irrespective of absolute mirror position. 

\begin{figure}
	\includegraphics[width=85mm]{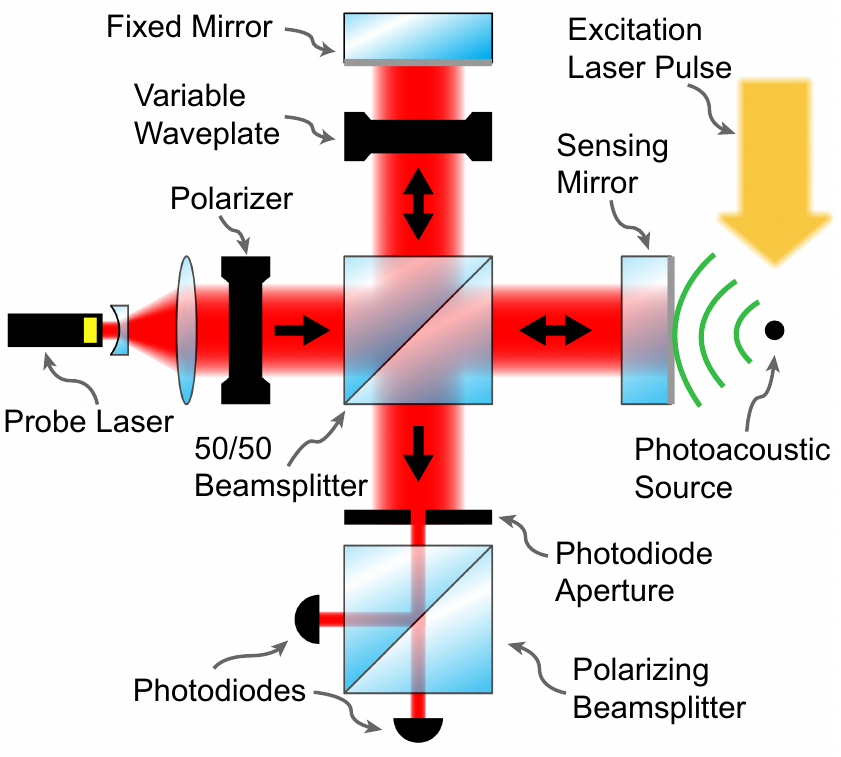}
	\caption{A schematic of the Michelson interferometer detector. Ultrasound passing through the sensing mirror changes the path length of the light, which alters the phase of the interference pattern. Quadrature phase detection allows high sensitivity to be achieved, whatever the mirror positions.\label{fig:Quadsetup}}
\end{figure}

A diagram of the detector setup can be seen in Fig. \ref{fig:Quadsetup}. The mirror position is  recovered from the detected intensities of the two polarizations, $I_{1,2}$ by first scaling them between $-1$ and $1$, then treating them as points on the unit circle: $\phi=\mathrm{atan2}(I_1,I_2)$.

\begin{figure}
	\includegraphics[width=85mm]{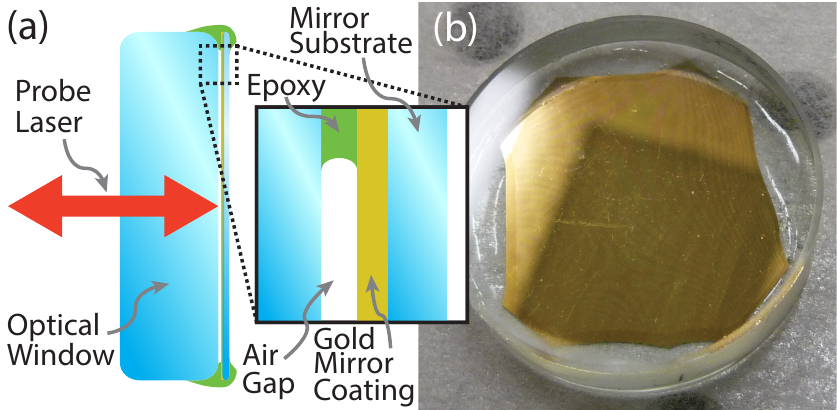}
	\caption{(a) Schematic of the sensing mirror showing probe laser. (b) The back side of the sensing mirror used in this experiment.\label{fig:real_mirror}}
\end{figure}

Our setup uses an expanded $5~\rm mW$, $633~\rm nm$ continuous wave helium-neon laser as the probe, and the detectors are amplified photodiodes with a $20~\rm MHz$ bandwidth. The photodiodes are apertured, which sets the spatial resolution of the system. The signal of each photodiode is recorded using a digital oscilloscope. The sensing mirror is a $150~\rm \mu m$ thick glass substrate with a gold reflective coating. To maintain mechanical stability, the mirror is mounted on an optical window by bonding it around the perimeter with epoxy resin. This method of bonding also ensures the presence of an air gap between the reflective surface and the window, eliminating the possibility of acoustic waves propagating into the window. The glass surface of the mirror is acoustically coupled to the medium being imaged, with the window becoming part of the interferometer arm. The resulting mirror can be seen in Fig. \ref{fig:real_mirror}. 

It is necessary to use such a thin mirror for two reasons. Firstly, a thick mirror suffers from the acoustic wave reflecting back and forth off the boundary of the substrate. This reflected wave interferes with the incoming wave, making the detected displacement useless. Multiple reflections are strongly suppressed in a mirror that is much thinner than the acoustic wavelength, so the displacement on the mirror surface accurately represents the incoming wave. Secondly, a mirror of thickness significantly greater than the acoustic wavelength is able to support surface waves. These surface, or \emph{Rayleigh} waves can exist whenever there is an impedance mismatch between two media, and are generated on the surface of the mirror when the photoacoustic pulse first reaches the boundary. The Rayleigh waves then propagate outwards across the surface and interfere with incoming photoacoustic waves.

To see the fundamental limitations of the detection system, it is useful to look at the theoretically achievable sensitivity. Like all Fabry-Perot and piezoelectric type detectors, the MI detector is sensitive to particle displacements, rather than directly to pressure. However because the MI detects the absolute position of a single plane, rather than the relative position of two planes, its sensitivity to pressure is easier to describe analytically. For a sinusoidal acoustic wave of pressure amplitude $p_0$, travelling in the positive $x$ direction, the corresponding particle displacement is given by:
\begin{equation}
	\xi(x,t)=\frac{p_0}{2\pi\nu z}\cos(kx-2\pi\nu t),
	\label{eqn:displacementfrequency}
\end{equation}
where $k$ is wavenumber, $\nu$ is frequency, and $z$ is the specific acoustic impedance of the propagation medium. For optical based pressure sensors the sensitivity, $S$ can be simply given as a proportion of optical intensity modulation per unit acoustic pressure: $S=1/I_{0}\frac{dI}{dp}$. Using the chain rule, this may be expanded to $\frac{dI}{dp}=\frac{dI}{d\phi}\frac{d\phi}{d\xi}\frac{d\xi}{dp}$. For the quadrature MI, $\frac{dI}{d\phi}$ is always at least $0.5I_0$, and $\frac{d\phi}{d\xi}$ is simply calculated from the expression for $\phi$. $\frac{d\xi}{dp}$ can be calculated from Eqn. \ref{eqn:displacementfrequency}, however it must be modified to describe the setup employed in our system. Firstly, since the pressure wave must propagate from the original medium into the glass mirror substrate, the pressure must be multiplied by the transmission coefficient: $T_p=2z_2/(z_1+z_2)$, where $z_{1,2}$ are the specific acoustic impedances of the first and second media respectively.\cite{Shutilov1988} Also, the mirror is essentially on a free boundary (since $z_{air}\ll z_{glass}$), so the particle displacement will be twice as great as in the bulk. Combining these terms gives the expression for the frequency dependent sensitivity:
 \begin{equation}
	S(\nu)=\frac{4n}{\nu\lambda(z_1+z_2)}.
	\label{eqn:sensitivity}
\end{equation}
Typical values of acoustic impedance for water and glass are $1.5\times10^6~\rm Pa~s~m^{-1}$ and $13.1\times10^6~\rm Pa~s~m^{-1}$ respectively. Taking $n = 1$ as the refractive index of air, and letting $\lambda = 633~\rm nm$, the sensitivity of our detector is $S(\nu)=0.43/\nu~\rm Hz~Pa^{-1}$. For a photoacoustic wave of frequency $1~\rm MHz$, the sensitivity of the detector should be 4.3\% optical modulation per $100~\rm kPa$ of peak acoustic pressure. This value can be compared directly with the sensitivity for an optical reflectance based detector of 0.19 to 0.81\% reported elsewhere in the literature.\cite{Paltauf1999} 

\begin{figure}
	\includegraphics[width=85mm]{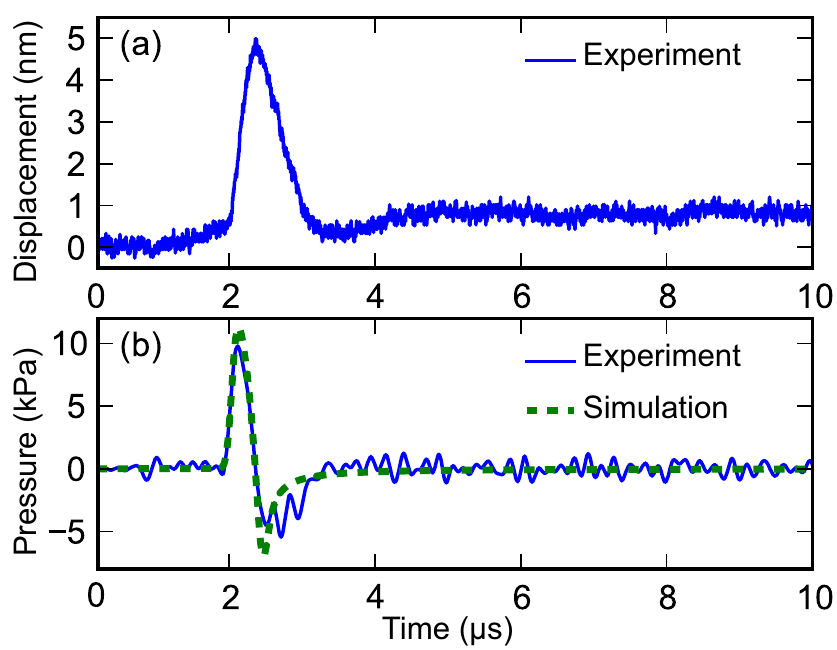}  
	\caption{(a) The detected mirror displacement due to a photoacoustic wave produced from a single source and (b), the corresponding pressure wave. A comparison between simulated and experimental pressure data shows the calculated sensitivity for the system is accurate.\label{fig:sim_result_compare}}
\end{figure}

Fig. \ref{fig:sim_result_compare} shows the actual detected displacement caused by a photoacoustic wave from a single source, positioned $3.3~\rm mm$ directly behind the detector. To recover the acoustic pressure from the displacement, the temporal derivative must be taken according to Eqn. \ref{eqn:pressure_displacement}, where $c=dx/dt$ has been used to change the variable of differentiation. A post processing low pass filter is applied to the detected displacement before the derivative is taken. This ensures the calculated pressure does not contain unphysical spikes which are artefacts of taking the derivative of a noisy signal.

The noise equivalent displacement in this trace is approximately $5~\rm\AA$, where the signal has been averaged for 64 pulses. The majority of the noise in this case is due to the amplified photodiodes, which may be improved using different equipment.

Fig. \ref{fig:sim_result_compare} also shows how the experimentally detected photoacoustic wave compares with a simulated one, and the agreement between the two support the values of sensitivity calculated previously. In the experiment, a long straight silicone tube with internal and external diameters of $0.5~\rm mm$ and $1.3~\rm mm$ respectively, was filled with diluted India ink with absorption coefficient $35~\rm cm^{-1}$ at $1064~\rm nm$. This value is similar to the optical absorption coefficient of blood at wavelengths commonly used in PAT. The tube was illuminated with a $10~\rm ns$ pulse of $1064~\rm nm$ light with fluence of $25~\rm mJ~cm^{-2}$. This fluence is well below the maximum permissible exposure for human skin of $100~\rm mJ~cm^{-2}$ based on American National Standards Institute recommendations.\cite{ANSI2007} The pulsed beam was collimated, and had a 1/e diameter of $1~\rm cm$. The simulation was performed using the method described by K\"ostli \emph{et al.}\cite{Kostli2001}, where Fig. \ref{fig:sim_result_compare}b shows a one-dimensional (1D) slice of the full 3D simulation. The parameters used in the simulation were the same as described for the experiment, however the simulation assumed an acoustically homogeneous propagation medium, so any acoustic effects of the silicone tube were ignored. The large diameter excitation beam minimized intensity variation across the tube, though due to the Gaussian nature of the beam, the stated fluence at the sample is only accurate to within 15\%. The fluence used in the simulation was adjusted within this range until the simulated amplitude most closely matched the detected one. As such, the calculated sensitivity is also only accurate to within this margin.

For the experimental measurement, the tube was submerged in a water filled glass cell. One wall of the cell was made of acoustically transparent polyethylene film, which was coupled to the sensing mirror of the interferometer with commercial ultrasound coupling gel. The detected pressure represents the pressure in the bulk of the glass substrate. This value was divided by the pressure transmission coefficient to give the acoustic pressure in the water, allowing a direct comparison with simulation.

\begin{figure}
	\includegraphics[width=85mm]{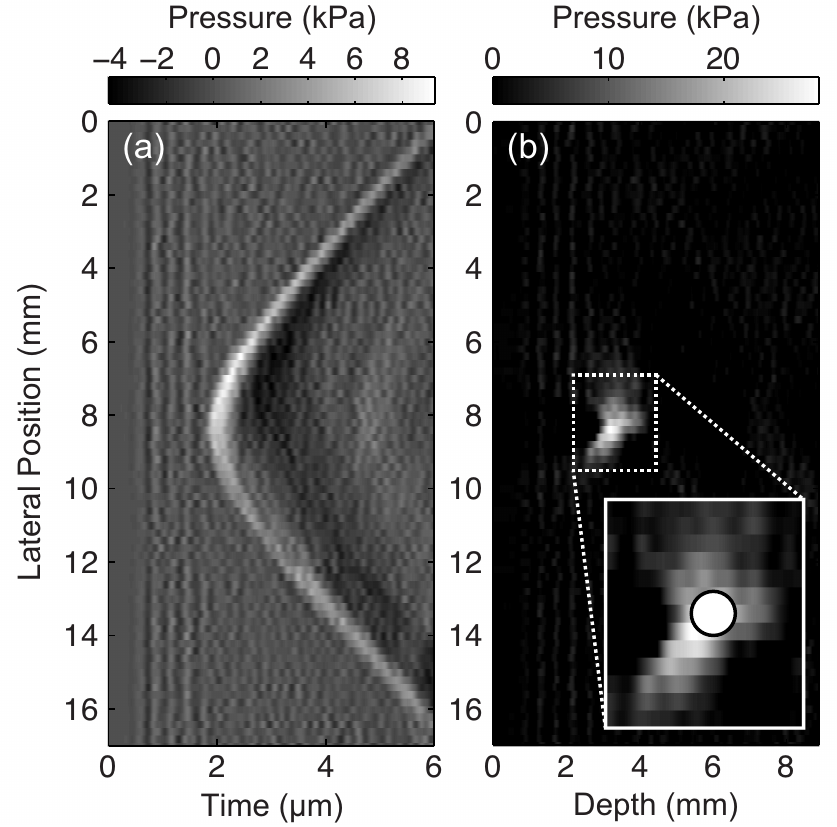}
	\caption{(a) Detected acoustic pressure from a single photoacoustic source and (b), the corresponding reconstructed image of the source. The size and location of the source is indicated by the circle in the magnified inset.\label{fig:XT_recon}}
\end{figure}

To create photoacoustic images, the photodiodes could be moved laterally to build up 1D or 2D scans. However the lower laser intensity at the edges of the expanded probe beam meant that data collected in these regions had a lower signal to noise ratio (SNR). Instead, the sample itself was scanned laterally. This allowed different regions of the generated ultrasonic wavefield to be sampled in an equivalent manner to scanning the photodiodes, while maintaining a high SNR. A 1D scan of the wavefield produced by the cylindrical optical absorber is shown in Fig. \ref{fig:XT_recon}. The source was the same India ink filled silicone tube as previously described, illuminated by the same pulsed laser. The photodiode aperture was set to a diameter of $200~\rm\mu m$, which was the same as the lateral step size.

The reconstruction shown in Fig. \ref{fig:XT_recon}b was performed using the \emph{kspaceLineRecon} function of the \emph{k-Wave} photoacoustic package.\cite{Treeby2010} The algorithm is based on Fourier transforms, and it is theoretically exact if pressure is detected over an infinite plane for infinite time.\cite{Kostli2001} In the reconstruction, the source has been positioned correctly, but has suffered some blurring and distortion which is consistent with other implementations of this inversion technique.\cite{Paltauf2007} The blurring is unsurprising given that the diameter of the source was only $500~\rm\mu m$ and the spatial separation of each data point was $200~\rm\mu m$.

The resolution of our system is currently limited by the size of the aperture in front of the photodiodes, which in turn is limited by the need to get sufficient laser power to the photodiodes. This is easily improved by increasing the laser power, and could be achieved using inexpensive diode lasers. Using a shorter wavelength probe laser would also be a simple way to boost sensitivity according to Eqn. \ref{eqn:sensitivity}. 

Manually scanning the photodetectors (or the source) to build up an image is too slow for real-time imaging applications, so any useful system must ultimately be capable of performing simultaneous detection over the whole surface. This could be simply achieved in our system by replacing the photodiode and aperture arrangement with fast gated intensified charge-coupled device cameras. CCDs detect intensity at many points across their surface simultaneously, eliminating the need to move the sample or the detector. The potential to use CCDs is a significant advantage over other proposed optical photoacoustic detectors, which have no easy route to simultaneous measurement of all elements over a large surface.

The current configuration of our sensing mirror (constructed from a thin glass substrate) may cause limitations to the detectable bandwidth needed in higher resolution systems, due to the possible reappearance of Rayleigh waves at higher frequencies. However this could be addressed by using polymer substrates impedance matched to water. Also, where deeper imaging is required, the bandwidth requirements of the detector are much more forgiving, since very high frequency acoustic waves are strongly attenuated in tissue. As such, the current sensing mirror is suitable for imaging depths beyond a few millimeters with no reduction in attainable resolution.
 
In summary, we have demonstrated the use of a Michelson interferometer as a photoacoustic detector. Quadrature phase detection removes the need to tune the sensitivity of the interferometer, allowing for the possibility of simultaneous full surface detection. We used the detector to produce proof-of-concept photoacoustic images with sub-millimeter resolution, and suggested ways that resolution could be improved. Future work will involve demonstrating simultaneous full surface measurement, with the aim of producing real-time 3D photoacoustic visualizations of the sub-cranial blood vessel vasculature. This would allow stroke researchers to visualize the dynamics of reperfusion in small animal subjects after inducing a stroke, in both high resolution and on a usefully short timescale.

Please cite the Applied Physics Letters version of this article, available online at: http://link.aip.org/link/doi/10.1063/1.4816427

\end{document}